\documentclass{wnna}
\usepackage{natbib}\usepackage{txfonts}\usepackage{balance}
\usepackage{graphicx}
\usepackage[a4paper]{hyperref}
\idline {1}{105}

\begin{document}
   
   \title{Super-AGB stars: evolution and associated nucleosynthesis}

\author{
M.L. \,Pumo\inst{1} 
\and L. \,Siess\inst{2}
\and R.A. \,Zappal\`a\inst{3}
}

\authoremail{mlpumo@ct.astro.it}

\institute{
INAF - Osservatorio Astrofisico di Catania,
Via S. Sofia 78, I-95123 Catania, Italy
\and
IAA - ULB, Boulevard du triomphe    
CP 226, B-1050 Bruxelles, Belgium
\and
Universit\`a di Catania,
Dip. di Fisica e Astronomia, Via S. Sofia 78,
I-95123 Catania, Italy
}

\authorrunning{Pumo et al.}

\titlerunning{Super-AGB stars}
   \subtitle{}

\abstract{ Based on evolutionary computations of 90 stellar models, we have analysed the impact of initial composition and core overshooting on the post-He-burning evolution and the associated nucleosynthesis of Super-AGB stars, pointing particular attention on the C-burning phase. Moreover the possible link between the transition masses $M_{up}$, $M_{N}$ and $M_{mas}$ (defined as the critical initial mass above which C-burning ignites, the minimum initial mass for an electron-capture supernova and the minimum initial mass for the completion of all the nuclear burning phases respectively) and the properties of the core during the core He-burning phase is also briefly discussed.
%in the light of these computations, the effects of initial composition and core overshooting on the values of $M_{up}$ (critical initial mass above which C-burning ignites) and $M_{mas}$ (minimum initial mass for the completion of all the nuclear burning phases) are presented. The possible link between the transition masses ($M_{up}$ and $M_{mas}$) and the properties of the core during the core He-burning phase is also briefly discussed.
 }

\maketitle

%________________________________________________________________

\section{Introduction}

In the stellar evolution context it is accepted the existence of two critical initial masses referred to as $M_{up}$ and $M_{mas}$. The former is defined as the critical initial mass above which C-burning ignites, while the latter indicates the minimum initial mass for the completion of all the nuclear burning phases \citep[e.g.][] {woosley02}. So-called Super-AGB (SAGB) stars fill the gap between $M_{up}$ and $M_{mas}$ \citep[e.g.][]{vienna06}. Thus these stars are massive enough to ignite carbon but, being unable to activate further nuclear burning stages, they end their life either forming a neon-oxygen white dwarf (NeO WD) or going through an electron-capture supernova (EC SN) becoming a neutron star \citep[e.g.][]{sp06}. The critical mass delineating the transition between these two fates is referred to as $M_{N}$. 
Although the general features of SAGB stars evolution seem to be well-established, open questions and uncertainties remain. In particular the advanced phases (post-He-burning evolution and final fate) of SAGB stars has not been studied in sufficient detail and the role of initial composition and other factors such as mass loss rate, internal mixing and effects of a companion star is still not well understood \citep[e.g.][]{h05,woosley02,p04}. In this work we address some of the above mentioned issues, pointing particular attention on the role of initial composition and core overshooting on the C-burning (CB) phase. Emphasis is also placed upon the possible link between the transition masses ($M_{up}$, $M_{N}$ and $M_{mas}$) and the properties of the core during the He-burning (HeB) phase.

%__________________________________________________________________

\section{Grids of stellar models}

Our work is based on the analysis of two grids of stellar models described in \citet{sp06}. We briefly report the main features of these evolutionary sequences in the following: $(i)$ the first grid consists of $70$ models without overshooting having initial masses ($M_{ini}$) between $7$ and $13M_{\odot}$ with $7$ different values of the initial metallicity (Z) in the range $10^{-5}$ to $0.04$; $(ii)$ the second grid consists of $20$ models with overshooting having $5\leq M_{ini}/M_{\odot} \leq 10.5$ for $Z= 10^{-4}$ and $0.02$; $(iii)$ starting from the pre-main sequence, all the models have been calculated by the code STAREVOL in the version described in \citet{s06a} with the differences reported in \citet{sp06}. 

%__________________________________________________________________

\section{Results and discussion}

\subsection{C-burning phase}

After the HeB phase, the temperature maximum increases as a consequence of core contraction and can move outward due to neutrino energy losses \citep[e.g.][]{sp06}, leading to an off-centre carbon ignition when the peak temperature reaches $\sim 6.5-7\cdot 10^8 K$ \citep[e.g.][]{thesis}. We find that the location of off-centre carbon ignition ($m_C$) is a decreasing function of $M_{ini}$ for a given mixing treatment and $Z$, but it increases with $Z$ when setting $M_{ini}$ and the mixing treatment, the only exceptions to this last behaviour being for $Z=10^{-5}$ and $0.04$. The decrease of $m_C$ with $M_{ini}$ is the consequence of the decreasing core degeneracy as $M_{ini}$ increases \citep{s06a}, while the behaviour of $m_C$ with $Z$ is essentially linked to opacity effects \citep[see][for details]{pumo07}. Once carbon is ignited, the evolution proceeds in two steps in all our models as found by \citet{s06a} for $Z=0.02$ models: first a carbon convective flash which induces a structural readjustment leading to a temporary quenching of the convective instability, and then the development of a convective flame that propagates to the stellar centre. However during the CB phase the exact evolutionary properties (e.g. maximum carbon luminosity during the flash and the flame, the mass covered by the convective instability during the flash and the duration of the flash and the flame) depend on $Z$, $M_{ini}$ and mixing treatment \citep[see][for details]{thesis}. Moreover the off-centre carbon ignition is so close to the centre ($m_C < 0.05M_{\odot}$) in some of the most massive members of our models (11$M_{\odot}$ model with $Z=0.008$, 11.3$M_{\odot}$ model with $Z=0.02$, 11$M_{\odot}$ model with $Z=0.04$ and $Z=0.02$ models with overshooting having $M_{ini}=9.0$ and $9.1M_{\odot}$) that only one convective episode occurs and there is no flame. 

\begin{figure}[h]
\resizebox{\hsize}{!}{\includegraphics[clip=true]{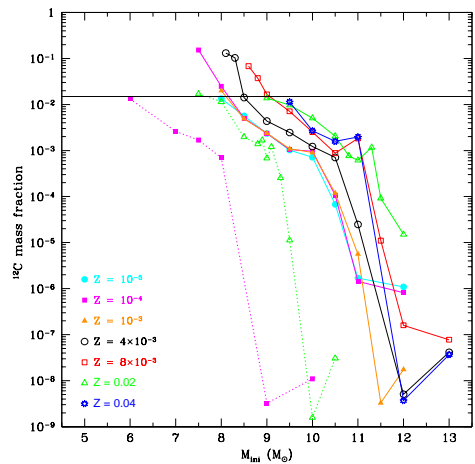}}
\caption{ \footnotesize Central mass fraction of $\rm^{12} C$ at the 
   end of the CB phase as a function of $M_{ini}$ for all our models. 
   The dotted lines refer to models with overshooting. The horizontal line indicates 
   the critical value ($\rm^{12} C=1.5\%$ by mass) for an explosion in case of EC SN event.} 
\label{fig:C12}
\end{figure}

In all our models the nucleosynthesis during the CB phase lead to a core made mainly of $\rm\,^{16} O$ ($\sim50-70\%$) and $\rm\,^{20} Ne$ ($\sim15-35\%$) with some traces of $\rm^{23} Na$ ($\sim 3-5\%$), $\rm^{24} Mg$ ($\sim 2-3\%$), $\rm^{25} Mg$ ($\sim 0.3-2\%$), $\rm^{26} Mg$ ($\sim 0.2-1\%$), $\rm\,^{22} Ne$ ($\sim 0.1-1\%$), $\rm\,^{21} Ne$ ($\sim 0.3-0.5\%$), $\rm\,^{27} Al$ ($\sim 0.1-0.8\%$) and $\rm\,^{17} O$ ($\sim 0.001-0.008\%$). It is to note that the $\rm^{23} Na$ is the most abundant nuclide among the minor components of the core. Thus one has a so-called NeONa core where the amount of $\rm^{24} Mg$ is too small (a factor $\sim 10$ lower than the required threshold value) to produce an explosion and, consequently, a disruption of the SAGB star in case of EC SN event \citep[see][and references therein for details on the explosion mechanism triggered by $\rm^{24} Mg$]{Guti05}. However, due to the $\rm^{12} C$ left unburned in the core after the CB phase, an explosion with subsequent disruption of the star could happen in all our models having $\rm^{12} C$ mass fraction above the horizontal line in Fig. \ref{fig:C12}, if the conditions for the EC SN are reached. In fact, according to the numerical simulations by \citet{Guti05}, small amounts of $\rm^{12} C$ ($\simeq 1.5\%$ by mass) are enough to completely disrupt the star in case of EC SN event.

%__________________________________________________________________________
\subsection{The critical He-free core masses}
\label{sez:core}

\begin{table*}
\caption{Values of $M_{up}$, $M_{N}$ and $M_{mas}$ (in solar units) taken from \citet{thesis} as a function of $Z$, for the models without (upper panel) and with (lower panel) overshooting respectively \citep[see also][]{sait07}. $M_{N}$ is obtained for different $\zeta$ values (namely $\zeta=-35$, $-70$, $-350$, $-700$ and $-4000$), where $\zeta$ is defined as the ratio of mass loss rate to core growth rate after the CB phase \citep[see][for details]{sait07}.}
\label{Table:Mass}
\begin{center}
\begin{tabular}{l|cccccccc}
 
   $Z$  & $M_{up}$ & \multicolumn{5}{c}{$M_{N}$}                              & $M_{mas}$\\
        &          & $^{-35}$ & $^{-70}$ & $^{-350}$ & $^{-700}$ & $^{-4000}$ &          \\
   \hline
   %  Z        Mup    Mn 1   Mn 2    Mn 3    Mn 4   Mn 5    Mmas
     0.00001 & 7.75 & 8.36 & 9.06  & 9.58  & 9.66  & 9.74  & 9.75  \\ 
      0.0001 & 7.25 & 8.28 & 9.02  & 9.53  & 9.59  & 9.65  & 9.66  \\ 
       0.001 & 7.75 & 8.33 & 9.06  & 9.60  & 9.66  & 9.71  & 9.72  \\ 
       0.004 & 8.05 & 8.75 & 9.48  & 10.13 & 10.19 & 10.25 & 10.26 \\ 
       0.008 & 8.45 & 9.39 & 9.99  & 10.52 & 10.59 & 10.65 & 10.66 \\ 
        0.02 & 8.90 & 9.65 & 10.44 & 10.83 & 10.85 & 10.90 & 10.93 \\ 
        0.04 & 9.25 & 9.66 & 10.20 & 10.77 & 10.83 & 10.88 & 10.89 \\ 
   \hline	      
      0.0001 & 5.75 & 6.77 & 7.20  & 7.55  & 7.59  & 7.62  & 7.63  \\ 
        0.02 & 7.25 & 7.89 & 8.39  & 8.77  & 8.81  & 8.83  & 8.83  \\ 
\end{tabular} 
\end{center}
\end{table*}
%---------------------

\begin{table*}
\caption{$M_{up}^{He-f}$, $M_{N}^{He-f}$ and $M_{mas}^{He-f}$ (in solar units) as a function of $Z$ for the models without (upper panel) and with (lower panel) overshooting respectively. $M_{N}^{He-f}$ is estimated for $\zeta=-35$, $-70$, $-350$, $-700$ and $-4000$.}
\label{tab:mCOcrit}
\begin{center}
\begin{tabular}{l|c  c c c c c  c} 
   $Z$  & $M_{up}^{He-f}$ & \multicolumn{5}{c}{$M_{N}^{He-f}$} &  $M_{mas}^{He-f}$\\
        &        &  $^{-35}$   & $^{-70}$ & $^{-350}$ & $^{-700}$  & $^{-4000}$ &  \\
   \hline
   %  Z         Mup    Mn 1    Mn 2   Mn 3   Mn 4    Mn 5   Mmas 
     0.00001 &  0.82 &  0.90 & 1.04 & 1.16 & 1.17 & 1.18 & 1.18\\ 
      0.0001 &  0.75 &  0.93 & 1.03 & 1.15 & 1.16 & 1.18 & 1.18\\ 
       0.001 &  0.82 &  0.93 & 1.05 & 1.15 & 1.16 & 1.17 & 1.18\\ 
       0.004 &  0.78 &  0.93 & 1.05 & 1.14 & 1.16 & 1.17 & 1.18\\ 
       0.008 &  0.80 &  0.89 & 1.04 & 1.14 & 1.16 & 1.17 & 1.17\\ 
        0.02 &  0.78 &  0.89 & 1.03 & 1.15 & 1.18 & 1.19 & 1.20\\ 
        0.04 &  0.81 &  0.90 & 1.04 & 1.15 & 1.17 & 1.18 & 1.18\\ 
    \hline		
      0.0001 &  0.87 &  1.04 & 1.14 & 1.21 & 1.22 & 1.23 & 1.23\\ 
        0.02 &  0.91 &  1.03 & 1.15 & 1.23 & 1.24 & 1.25 & 1.26\\ 
\end{tabular}
\end{center}
\end{table*}
%-----------------

As outlined in the introduction, SAGB stars are at the cross road of stellar evolution. So a study on these stars is also able to improve our knowledge about the critical masses that set the transition among various groups of stars characterised by a different evolutionary path and final fate. In this framework, since the evolutionary properties during the HeB phase strongly affect the subsequent evolution of the stars \citep[e.g.][]{sp06,woosley02}, we have tried to make out if the origin of the transition masses $M_{up}$, $M_{N}$ and $M_{mas}$ could be traced back to the properties of the core during the HeB phase and, in particular, to the He-free core mass (i.e. CO core mass) at the end of HeB phase.
 
\begin{figure}[ht]
\resizebox{\hsize}{!}{\includegraphics[clip=true]{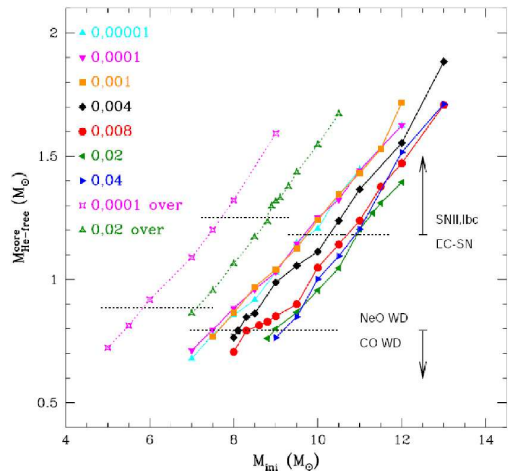}}
\caption{ \footnotesize CO core mass (see note \ref{note:COcore}) at the end of HeB phase as a function of $M_{ini}$ for all our models. The dotted horizontal lines roughly delineate the transition among different outcomes of stellar evolution.} \label{fig:hefree}
\end{figure}

Following this idea, from the data shown in Fig. \ref{fig:hefree}, we have estimated the CO core mass\footnote{In our calculations the CO core mass is evaluated considering the mass coordinate of the locus of maximum nuclear energy production due to the HeB as upper edge of such core. \label{note:COcore}} at the end of core HeB (when the central He mass fraction has dropped below $10^{-4}$) for stars with $M_{ini}$ mass equal to $M_{up}$, $M_{N}$ and $M_{mas}$ (see Table \ref{Table:Mass}). These values, referred to as $M_{up}^{He-f}$, $M_{N}^{He-f}$ and $M_{mas}^{He-f}$ respectively, represent the critical CO core masses above which the nature of the end-product of stellar evolution changes. From our results (see Table \ref{tab:mCOcrit}) it appears that the critical CO core masses are weakly sensitive to the initial composition for a given mixing treatment. Stars with $M_{ini}=M_{up}$ have on average $M_{up}^{He-f}$ equal to $\sim 0.79M_{\odot}$ for models without overshooting and equal to $\sim 0.89M_{\odot}$ for models with overshooting. Therefore, stars developing CO core mass less than $\sim 0.79M_{\odot}$ (or $\sim 0.89M_{\odot}$ for models with overshooting) will never ignite carbon, regardless of the $Z$ value. Concerning the core mass limit above which an iron core forms, we find that on average $M_{mas}^{He-f}$ is equal to $\sim 1.18M_{\odot}$ for models without overshooting and $\sim 1.24M_{\odot}$ for models with overshooting. Depending on the prescriptions for the mass loss and core growth rates (i.e. according to the $\zeta$ values), the value of $M_{N}^{He-f}$ changes and can cover a mass range of $\sim0.35M_{\odot}$. However, as pointed out in \citet{sait07}, the formation of a NeO WD or a neutron star seems to be connected more intimately to the second dredge-up phenomenon rather than to the exact value of $M_{N}^{He-f}$. In fact this phenomenon is able to hamper a possible EC SN event by reducing the core mass below the Chandrasekhar mass limit \citep{vienna06}. 

In the light of results, we believe it is worthwhile to point our attention on the nucleosynthesis during the post-CB phases, in particular investigating the possibility of s-nuclei nucleosynthesis during the thermally pulsing SAGB phase.

%___________________________________________________________
\bibliographystyle{aa}

\end{document}